\documentclass[pre, showpacs, showkeys, nofootinbib]{revtex4}
\usepackage{latexsym}
\usepackage{amsmath}
\usepackage[dvips]{graphicx}

\begin {document}
\title{Lifetime of small systems controlled by autocatalytic reactions}
\author{L. \surname{P\'al}}
\email[Electronic address:]{lpal@rmki.kfki.hu} \affiliation{KFKI
Atomic Energy Research Institute H-1525 Budapest 114, POB 49
Hungary}

\begin{abstract}
By using the point model of reaction kinetics we have studied the
stochastic properties of the lifetime of small systems controlled
by autocatalytic reaction $A + X \rightleftharpoons X + X, \;\; X
\rightarrow B$. Assuming that a system is living only when the
number of autocatalytic particles is larger than zero but smaller
than a positive integer $N$, we have calculated the probability of
the lifetime provided that the number of substrate particles $A$
is kept constant by a suitable reservoir, and the end-products $B$
do not take part in the reaction. We have shown that the density
function of the lifetime is strongly asymmetric and in certain
cases it has a well-defined minimum at the beginning of the
process. It has been also proven that the extinction probability
of systems of this type is exactly $1$.
\end{abstract}



\maketitle

\section{Introduction}

Finding models for calculations of lifetime probability of systems
evolving in accordance to well-defined rules seems to be a very
attractive problem \cite{kendall50, bharucha53, eigen71, segre98,
ferreira03} not only in physics but also in many other fields of
sciences. The system evolution is usually controlled by stochastic
transformations of components which are called in the sequel
"particles". In the present model the transformation is realized
by autocatalytic reactions which play important role in
organization and coordination of systems having small spatial
volume and containing few particles only.

In this paper we are dealing with small systems which are in
living state when the number of autocatalytic particles remains
within the interval $(0,N)$, where $N$ is a positive integer,
while the number of substrate particles is kept constant by a
suitable reservoir. One can say that both the absence and the
large "dose" of $X$ particles lead to the extinction of the
system. We would like to determine the lifetime probability of
such systems.

We choose from many possible autocatalytic reactions one of the
simplest, namely $A + X \rightleftharpoons X + X,\;\; X
\rightarrow B$, where $A$ denotes the substrate particles the
number of which
\begin{figure} [ht!] \protect \centering{
\includegraphics[height=1.6cm, width=8cm]{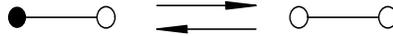}}\protect
\vskip -0.3cm \protect \caption{\label{fig1}{\footnotesize
Illustration of a reversible autocatalytic reaction by a simple
graph.}}
\end{figure}
is assumed to be constant, and $X$ symbolizes the autocatalytic
particles. The end-product particles $B$ do not take part in the
reaction. For the sake of descriptiveness  the reaction $A + X
\rightleftharpoons X + X$  is illustrated by a simple graph shown
in FIG. \ref{fig1}, where the white circle corresponds to an
autocatalytic while the black one to a substrate particle.

\begin{figure} [ht!] \protect \centering{
\includegraphics[height=2.5cm, width=8cm]{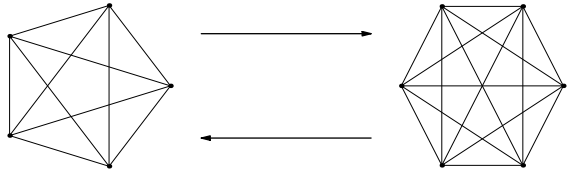}}\protect
\vskip 0.2cm \protect \caption{\label{fig2}{\footnotesize
Illustration of a system transformation caused by autocatalytic
reaction.}}
\end{figure}

A system consisting of $N_{A}$ substrate and $N_{X}$ autocatalytic
particles can be represented by a set of complete graphs $K_{N}$
on $V=N_{A} + N_{X}$ points, where $N_{A}$ is fixed and $\max
N_{X}$ is denoted by $N$. A randomly chosen point of a graph may
be either $A$ or $X$, the only requirement is that the total
numbers of $A$ and $X$ points must be $N_{A}$ and $N_{X}$,
respectively. Clearly, the number of complete graphs
characterizing a given systems is given by $1/B(N_{A}+1,
N_{X}+1)$, where $B(x, y)$ is the beta function. Since every two
points adjacent, the possibility of the interaction between any
two particles of the system is the same. Therefore, it is obvious
that the point model of reaction kinetics introduced by
\cite{lpal04} can be applied. As an example, FIG. \ref{fig2}
illustrates a transformation which brings about a system of $V=6$
particles from a system of $V=5$ particles and vice versa.

In Section II we derive equations determining the probability
$p(t,n)$ of finding $n$ autocatalytic particles at the time moment
$t$ in a system of volume $V$ provided that at $t=0$ the number of
particles was $i$. In Section III we show that the extinction
probability of the system is $1$, while in Section IV results of
exact calculations for $N=3$ and $N=5$ are presented. The main
conclusions are summarized in Section V.

\section{Systems with limited number of autocatalytic
particles}

Now, we formulate the basic equation. Let $\xi(t)$ be the number
of autocatalytic particles $X$ at time instant $t \geq 0$ in a
system open for substrate particles $A$. In the sequel, we would
like to define a special system existing only when $\xi(t)$
remains within the interval $(0, N)$, where $N$ is a positive
integer. We say the system is in \textit{the state} ${\mathcal
S}_{n}$, if $\xi(t)=n$. Obviously, a system will be annihilated
when it enters into the state either ${\mathcal S}_{0}$ or
${\mathcal S}_{N}$. For the sake of simplicity we call a system of
this type \texttt{SL} system.

In order to characterize the stochastic behavior of an \texttt{SL}
system we define the transition probabilities~\footnote{In order
to simplify the notations the index referring to the initial state
${\mathcal S}_{i}$ will be omitted where it does not cause
confusions.}
\[ {\mathcal P}\{\xi(t)=n|\xi(0)=i\} = p_{i,n}(t) = p_{n}(t),\;\;\;\;\;\;
n = 0, 1, 2, \ldots, N, \] which are determined by the following
equations:
\begin{equation} \label{1}
\frac{dp_{0}(t)}{dt} = \gamma p_{1}(t),
\end{equation}
\begin{equation} \label{2}
\frac{p_{n}(t)}{dt} = -n(N_{A} \alpha +n\beta - \beta + \gamma)
p_{n}(t) + (n-1)N_{A} \alpha p_{n-1}(t) + (n+1)(n\beta + \gamma)
p_{n+1}(t),
\end{equation}
\[ n = 1, 2, \ldots, N-2. \]
Since there is no way out from the state ${\mathcal S}_{N}$ the
equation for $p_{N-1}(t)$ should have the form
\begin{equation} \label{3}
\frac{p_{N-1}(t)}{dt} = -(N-1)(N_{A} \alpha + N\beta - 2\beta +
\gamma) p_{N-1}(t) + (N-2)N_{A} \alpha p_{N-2}(t),
\end{equation}
and, finally the equation
\begin{equation} \label{4}
\frac{p_{N}(t)}{dt} = (N-1)N_{A} \alpha p_{N-1}(t)
\end{equation}
determines the probability of the transition ${\mathcal S}_{i}
\rightarrow {\mathcal S}_{N}$ during the time $t \geq 0$.
Introducing the vector
\[\vec{\bf p}(t, N-1) = \left( \begin{array}{c} p_{N-1}(t) \\ p_{2N-2}(t) \\
\vdots \\ p_{1}(t) \end{array} \right)\] and the matrix
\begin{equation} \label{5}
{\bf A}_{N-1} = \begin{pmatrix} - D_{N-1} & B_{N-1} & 0 & 0 & \cdots & 0 & 0 & 0 \\
C_{N-2} & - D_{N-2} & B_{N-2} & 0 & \cdots & 0 & 0 & 0 \\
0 & C_{N-3} & - D_{N-3} & B_{N-3} & \cdots & 0 & 0 & 0 \\
\vdots & \vdots & \vdots & \vdots & \cdots & \vdots & \vdots & \vdots \\
0 & 0 & 0 & 0 & \cdots & C_{2} & - D_{2} & B_{2} \\
0 & 0 & 0 & 0 & \cdots & 0 &  C_{1} & - D_{1}
\end{pmatrix},
\end{equation}
we can write Eqs. (\ref{2}) and (\ref{3}) in the following compact
form:
\begin{equation} \label{6}
\frac{d\vec{\bf p}(t,N-1)}{dt} = \beta {\bf A}_{N-1}\;\vec{\bf
p}(t,N-1),
\end{equation}
with initial conditions $p_{n}(0) = \delta_{n,i} \;\; \text{for}
\;\; 1 \leq n \leq N-1$. The elements of the matrix
$\mathbf{A}_{N-1}$ are given by
\begin{equation} \label{7}
B_{n} = (n-1)a , \;\;\;\;\;\; D_{n} = n(a+b+n-1), \;\;\;\;\;\;
\mbox{and} \;\;\;\;\; C_{n} = (n+1)(b+n),
\end{equation}
\[ 1 \leq n \leq N-1, \]  where
\begin{equation} \label{8}
a = N_A\;\frac{\alpha}{\beta}, \;\;\;\;\;\; \text{and}
\;\;\;\;\;\; b = \frac{\gamma}{\beta}.
\end{equation}
One has to note that for the complete description of the process
we need also the equations (\ref{1}) and (\ref{4}). The matrix
${\bf A}_{N-1}$ is a \textit{normal Jacobi matrix}. (See Ref.
\cite{dean56, karlin58, bellman60}.) It can be proven that the
\textit{eigenvalues of $\mathbf{A}_{N-1}$ are different, negative
real numbers}. The proof is given in Appendix A. Consequently we
obtain
\[ \lim_{t \rightarrow \infty} p_{n}(t) = 0, \;\;\;\;\;\;
\mbox{if} \;\;\;\;\;\; n = 1, 2, \ldots, N-1, \] but
\begin{equation} \label{9}
\lim_{t \rightarrow \infty} p_{0}(t) = w_{0}, \;\;\;\;\;\; \lim_{t
\rightarrow \infty} p_{N}(t) = w_{N} \;\;\;\;\;\; \mbox{and}
\;\;\;\;\;\; w_{0} + w_{N} = 1,
\end{equation}
i.e. if $t \rightarrow \infty$, then an \texttt{SL} system can be
found  either in ${\mathcal S}_{0}$ or in ${\mathcal S}_{N}$
state.

\section{Lifetime probability}

The random time $\theta_{i}, \; 1 \leq i \leq N-1$ due to the
transitions either ${\mathcal S}_{i} \rightarrow {\mathcal S}_{0}$
or ${\mathcal S}_{i} \rightarrow {\mathcal S}_{N}$ is called
\textit{the lifetime of the} \texttt{SL} \textit{system}. In other
words, the lifetime is a random time interval at the end of which
the number of autocatalytic particles becomes either zero or $N$.
Formulating more precisely we can write
\[ \{\theta_{i} \leq t\} = \{\xi(t)=0|\xi(0)=i\} \cup
\{\xi(t)=N|\xi(0)=i\}\] and since $\{\xi(t)=0|\xi(0)=i\} \cap
\{\xi(t)=N|\xi(0)=i\} = \emptyset$, we have
\begin{equation} \label{10}
{\mathcal P}\{\theta_{i} \leq t\} = H_{i}(t) = p_{i,0}(t) +
p_{i,N}(t).
\end{equation}
By using Eqs. (\ref{9}) we see immediately that
\begin{equation} \label{11}
\lim_{t \rightarrow \infty} H_{i}(t) = 1, \;\;\;\;\;\; \forall\;1
\leq i \leq N-1,
\end{equation}
i.e. the probability that the lifetime of the system is finite is
equal to $1$.~\footnote{This statement is equivalent to that the
probability of the extinction of the system is equal to $1$.}

\section{Systems with small number of states}

The mathematical properties of the equation (\ref{6}) are
well-known \cite{karlin57}. Here, we do not wish to make detailed
analysis of Eq. (\ref{6}) and do not derive its formal solution,
but instead we try to study the basic properties of exact
solutions in those cases when the maximal number of autocatalytic
particles is $N=3$ and $N=5$, i.e. the number of states is four
and six, respectively.

\subsection{Four states systems}

In order to have an insight into the dynamics of the process
producing random transitions between different states of the
system we are dealing now with a very simple system having only
four states, namely ${\mathcal S}_{0}, \;{\mathcal
S}_{1},\;{\mathcal S}_{2},$ and ${\mathcal S}_{3}$. In this case
the equations to be solved are
\begin{eqnarray}
\frac{dp_{0}(t)}{dt} & = & \gamma p_{1}(t), \label{12} \\
\frac{dp_{1}(t)}{dt} & = & - (\alpha\;N_{A} +
\gamma)\;p_{1}(t) + 2\;(\beta + \gamma)\;p_{2}(t), \label{13} \\
\frac{dp_{2}(t)}{dt} & = & \alpha\;N_{A}\;p_{1}(t) -
2\;(\alpha\;N_{A} + \beta + \gamma)\;p_{2}(t),
\label{14} \\
\frac{dp_{3}(t)}{dt} & = & 2\;\alpha\;N_{A}\;p_{2}(t). \label{15}
\end{eqnarray}
The initial conditions to be used are
$p_{0}(0)=p_{2}(0)=p_{3}(0)=0$ and $p_{1}(0)=1$. Introducing the
notations (\ref{8}), after elementary calculations one obtains
\begin{equation} \label{16}
p_{0}(t) = b\frac{d-a-b-2}{2d\; \omega_{1}}\;(1 -
\exp\{-\omega_{1} \beta t\}) + b\frac{d+a+b+2}{2d\;
\omega_{2}}\;(1 - \exp\{-\omega_{2} \beta t\}),
\end{equation}
\begin{equation} \label{17}
p_{1}(t) = \frac{\exp\{-\omega_{2} \beta t\} + \exp\{-\omega_{1}
\beta t\}}{2} + \frac{a+b+2}{2d}\;\left(\exp\{-\omega_{2} \beta
t\} - \exp\{-\omega_{1} \beta t\}\right),
\end{equation}
\begin{equation} \label{18}
p_{2}(t) = \frac{a}{d}\left(\exp\{-\omega_{2} \beta t\} -
\exp\{-\omega_{1} \beta t\}\right),
\end{equation}
and
\begin{equation} \label{19} p_{3}(t) =
2\;\frac{a^{2}}{d}\;\left(\frac{1 - \exp\{-\omega_{2} \beta
t\}}{\omega_{2}} - \frac{1 - \exp\{-\omega_{1} \beta
t\}}{\omega_{1}}\right),
\end{equation}
where
\begin{equation} \label{20}
d = \sqrt{a^{2}+(b+2)^{2}+2a(5b+6)}, \;\;\;\;\;\; \omega_{1} =
\frac{1}{2}(\lambda + d) \;\;\;\;\;\; \mbox{and} \;\;\;\;\;\;
\omega_{2} = \frac{1}{2}(\lambda - d),
\end{equation}
while
\begin{equation} \label{21}
\lambda = 2 + 3\;(a + b).
\end{equation}

\begin{figure} [ht!]
\protect \centering{
\includegraphics[height=12cm, width=8cm]{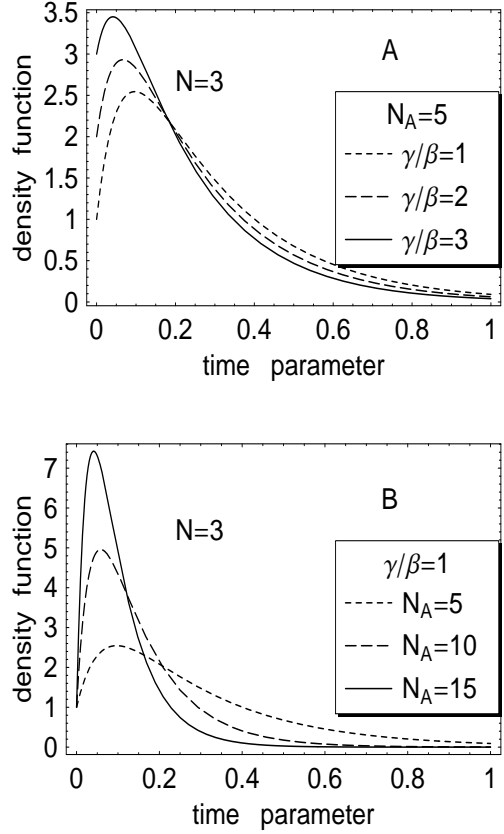}}\protect
\vskip 0.2cm \protect \caption{\label{fig3}{\footnotesize
Probability density functions of the system lifetime $\theta_{1}$.
Part A shows the density functions at different $\gamma/\beta$
values when $N_{A}=5$, while part B illustrates the effect of the
variation of the number of $A$ particles at fixed $\gamma/\beta$
value. The ratio $\alpha/\beta=1$.}}
\end{figure}

According to the theorem on eigenvalues nonnegative eigenvalue
cannot be occurred, therefore $\lambda > d$. Naturally, this
inequality can be simply proven for all positive $a$ and $b$ by
using Eqs. (\ref{20}) and (\ref{21}).~\footnote{Suppose that
$\lambda = d$, i.e.
\[ 3(a+b)^{2} - a^{2} + (b+2)^{2} + 2a(5b+6) +  2 = 0. \]
It is elementary to show the roots of this equation for $b$ to be
complex, if $0 < a < 0.4467$, and negative, if $a \geq 0.4467$.}

The probability density functions of the system lifetime
$\theta_{1}$ are plotted on FIG. \ref{fig3} at different values of
the number of $A$ particles (part $B$ of the figure) and at
different $\gamma/\beta$ ratios (part $A$ of the figure). It is
remarkable that the density functions are strongly asymmetric, and
the most probable lifetimes measured in $\beta^{-1}$ units are
found near the zero of the time parameter.

As seen from Eqs. (\ref{12})-(\ref{15}) we obtain the following
limit probabilities:
\begin{equation} \label{22}
\lim_{t \rightarrow \infty} p_{0}(t) = w_{0} = \frac{b\;(1 + a +
b)} {a^{2} + b\;(1 + a + b)}, \;\;\;\; \lim_{t \rightarrow \infty}
p_{3}(t) = w_{3} = \frac{a^{2}} {a^{2} + b\;(1 + a + b)},
\end{equation}
and
\[ \lim_{t \rightarrow \infty} p_{1}(t) =
\lim_{t \rightarrow \infty} p_{2}(t) = 0. \] According to this,
after elapsing sufficiently long (i.e. infinite) time, the system
can be found almost surly in "dead" state either with probability
$w_{0}$ in the state ${\mathcal S}_{0}$ or with probability
$w_{3}$ in the state ${\mathcal S}_{3}$. Obviously
$w_{0}+w_{3}=1$.

\subsection{Six states systems}

Now, we would like to see how the number of possible living states
effects on the properties of the lifetime. If the number of states
is six, i.e. $N=5$, then the system has four living and two dead
states. According to the notation introduced earlier, the living
states are ${\mathcal S}_{1}, {\mathcal S}_{2}, {\mathcal S}_{3},
{\mathcal S}_{4}$, while the dead ones ${\mathcal S}_{0}$ and
${\mathcal S}_{5}$.

\begin{figure} [ht!]
\protect \centering{
\includegraphics[height=4.6cm, width=7cm]{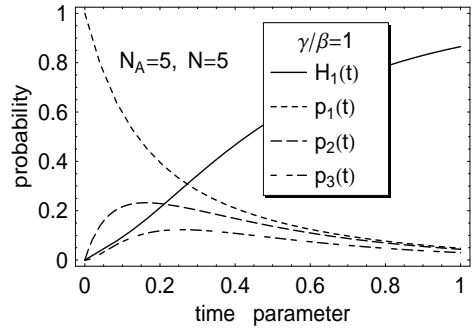}}\protect
\vskip 0.2cm \protect \caption{\label{fig4} {\footnotesize
Dependence of probabilities $p_{1}(t), p_{2}(t), p_{3}(t)$ and the
system lifetime distribution $H_{1}(t)$ on $t$ measured in
$\beta^{-1}$ units. The ratio $\alpha/\beta=1$.}}
\end{figure}

\begin{figure} [ht!]
\protect \centering{
\includegraphics[height=4.6cm, width=7cm]{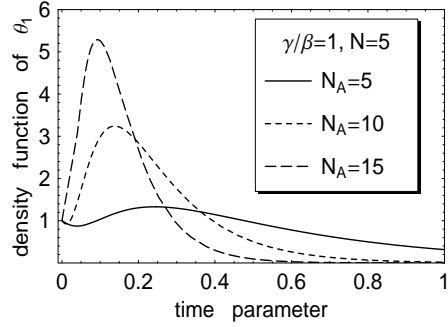}}\protect
\vskip 0.2cm \protect \caption{\label{fig5} {\footnotesize Density
function of the system lifetime at different values of the number
of $A$ particles at $\gamma/\beta=1$. The number of living states
is equal to $4$. The time $t$ is measured in $\beta^{-1}$ units.
The ratio $\alpha/\beta=1$.}}
\end{figure}

By using numerical method for solving the equation system
(\ref{1})-(\ref{4}) in the case of $N=5$ we obtain the
probabilities $p_{n}(t), \; n=0,1,\ldots,5$ vs. time parameter
curves shown some of them in FIG. \ref{fig4}. What remarkable is
the character of curves does not change with increasing number of
possible states. It can be proven that the higher the number of
possible states the larger is the probability that the system
lifetime is longer than $t$.

\begin{figure} [ht!]
\protect \centering{
\includegraphics[height=5.2cm, width=8cm]{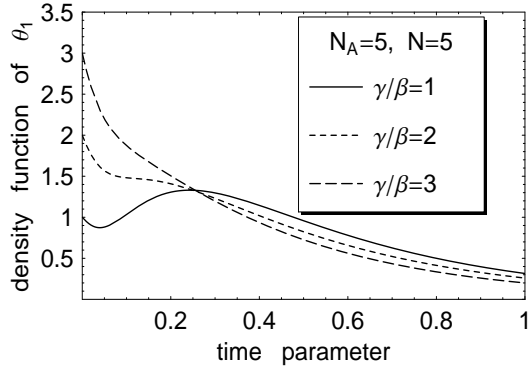}}\protect
\vskip 0.2cm \protect \caption{\label{fig6} {\footnotesize Density
function of the system lifetime at different values of the ratio
$\gamma/\beta$ and at the fixed number of $A$ particles. The
number of living states is $4$ and the time $t$ is measured in
$\beta^{-1}$ units. The ratio $\alpha/\beta=1$.}}
\end{figure}

\begin{figure} [ht!]
\protect \centering{
\includegraphics[height=5.2cm, width=8cm]{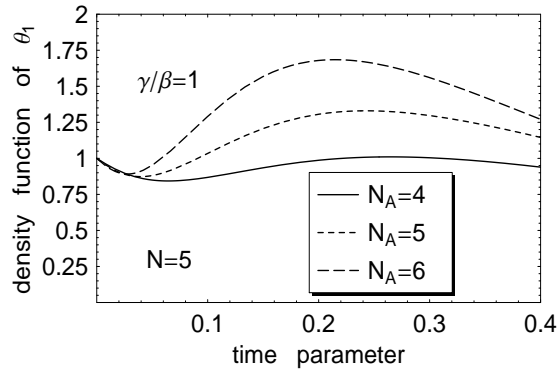}}\protect
\vskip 0.2cm \protect \caption{\label{fig7} {\footnotesize Density
function of the system lifetime at the beginning of the process
for three different values of the number of $A$ particles. The
number of living states is $4$, the time $t$ is measured in
$\beta^{-1}$ units and $\gamma/\beta=1$. The ratio
$\alpha/\beta=1$.}}
\end{figure}

Density functions of the system lifetime $\theta_{1}$ measured in
$\beta^{-1}$ units are seen in FIG.~\ref{fig5} at values $N_{A} =
5, 10, 15$. The calculations are performed at fixed
$\gamma/\beta=1$ ratio. One can see the curves are strongly
asymmetric and the most probable lifetime decreases with
increasing $N_{A}$. The reason for this is clear: the speed of the
reaction increases with the number of $A$ particles, and so if
$N_{A}$ is large, then the state ${\mathcal S}_{N}$ can be reached
in a time interval shorter than that is needed when $N_{A}$ is
small. It is remarkable that we see in FIG.~\ref{fig6} showing the
density function of $\theta_{1}$ at different $\gamma/\beta$
ratios. If the decay intensity $\gamma$ is approximately equal to
the intensity $\beta$ of the reverse reaction $X + X \rightarrow X
+ A$, then a well-defined minimum appears in the density vs.
lifetime curve at the very beginning of the process, because the
autocatalytic production of $X$ particles cannot compensate the
effect of the decay process. After elapsing some time the role of
the autocatalytic process becomes perceptible and, if the ratio
$\gamma/\beta$ is smaller than $3$, then the density function
reaches a maximum which is followed by a monotonously decreasing
part of the curve. The density function at the beginning of the
process is shown in FIG.~\ref{fig7} for three $N_{A}$ values.

\section{Conclusions}

Stochastic dynamics of small systema controlled by autocatalytic
reaction $X + X \rightleftharpoons X + A, \; X \rightarrow B$ is
studied. The dead state of systems is defined and the probability
distribution of the lifetime are calculated by exact means. The
density function of the lifetime is found to be strongly
asymmetric (skewed right). The most probable lifetime (mode of the
lifetime) decreases with increasing number of the substrate
particles $A$. If the self-decay intensity of the autocatalytic
particles is approximately equal to the intensity of the reverse
reaction $X + X \rightarrow X + A$, then a well-defined minimum
appears in the density function of the lifetime near the beginning
of the process. It is remarkable that the basic character of the
system's dynamics does not vary significantly with increasing
number of possible state.

\appendix

\section{Eigenvalues}

We would like to prove that the eigenvalues of the matrix
$\mathbf{A}_{N-1}$ are different, negative real numbers. Let us
introduce the matrix
\begin{equation} \label{a1}
\mathbf{J}_{n} = - \mathbf{A}_{n} = \begin{pmatrix} G_{n} &  H_{n} & 0 & 0 & \cdots & 0 & 0 & 0 \\
F_{n-1} &  G_{n-1} &  H_{n-1} & 0 & \cdots & 0 & 0 & 0 \\
0 &  F_{n-2} &  G_{n-2} &  H_{n-2} & \cdots & 0 & 0 & 0 \\
\vdots & \vdots & \vdots & \vdots & \cdots & \vdots & \vdots & \vdots \\
0 & 0 & 0 & 0 & \cdots &  F_{2} &  G_{2} &  H_{2} \\
0 & 0 & 0 & 0 & \cdots & 0 &   F_{1} &  G_{1}
\end{pmatrix},\;\;\;\;\;\; \text{where}
\;\;\;\;\;\; n \in {\mathcal Z}_{+},
\end{equation}
while
\begin{equation} \label{a2}
F_{j} = - (j+1)\;(b+j), \;\;\;\;\;\; G_{j} = j\;(a+b+j-1),
\;\;\;\;\;\; H_{j} = -(j-1)\;a,
\end{equation}
where $a$ and $b$ are positive real numbers given by Eq.
(\ref{8}). Since $F_{j}/H_{j-1} > 0$ there is a diagonal matrix
$\mathbf{S}_{n}$ such which transforms $\mathbf{J}_{n}$ into a
symmetric matrix
\begin{equation} \label{a3}
\mathbf{R}_{n} =
\mathbf{S}_{n}^{-1}\;\mathbf{J}_{n}\;\mathbf{S}_{n},
\end{equation}
the eigenvalues of which are exactly the same as those of
$\mathbf{J}_{n}$. The entries of $\mathbf{S}_{n}$ are
\[ s_{i,j} = \delta_{i,j}\; s_{j}, \;\;\;\;\;\; \text{where} \;\;\;\;\;\;
s_{j} = \sqrt{\frac{F_{j}}{H_{j-1}}}\;s_{j-1}, \;\;\;\;\;\;
\text{and} \;\;\;\;\;\; s_{1} = 1. \] It is easy to show that

{\footnotesize
\begin{equation} \label{a4}
\mathbf{R}_{n} = \begin{pmatrix} G_{n} &  \sqrt{H_{n}\;F_{n-1}} & 0 & 0 & \cdots & 0 & 0 & 0 \\
\sqrt{H_{n}\;F_{n-1}} &  G_{n-1} &  \sqrt{H_{n-1}\;F_{n-2}} & 0 & \cdots & 0 & 0 & 0 \\
0 &  \sqrt{H_{n-1}\;F_{n-2}} &  G_{n-2} &  \sqrt{F_{n-3}\;H_{n-2}} & \cdots & 0 & 0 & 0 \\
\vdots & \vdots & \vdots & \vdots & \cdots & \vdots & \vdots & \vdots \\
0 & 0 & 0 & 0 & \cdots &  \sqrt{F_{2}\;H_{3}} &  G_{2} &  \sqrt{F_{1}\;H_{2}} \\
0 & 0 & 0 & 0 & \cdots & 0 &  \sqrt{F_{1}\;H_{2}} &  G_{1}
\end{pmatrix},
\end{equation}}

\noindent and clearly $\mathbf{R}_{n}$ is a symmetric Jacobi
matrix. As known, if it is positive definite, then all of its
eigenvalues are different, real and positive. A symmetric matrix
is positive definite if the principal minors of the corresponding
determinant are all positive. Let us denote by
\begin{equation} \label{a5}
{\mathcal D}_{j} = \text{det}\;\vert\mathbf{R}_{j}\vert
\end{equation}
the determinant due to the matrix $\mathbf{R}_{j}$. Taking into
account that ${\mathcal D}_{0} = 1$ by definition, we have to show
that ${\mathcal D}_{j} > 0, \;\; \forall\; j=1,2,\ldots,n$. After
simple calculation we obtain
\begin{equation} \label{a6}
{\mathcal D}_{j} = G_{j}\;{\mathcal D}_{j-1} -
H_{j}\;F_{j-1}\;{\mathcal D}_{j-2},
\end{equation}
and by induction we have
\begin{equation}
{\mathcal D}_{j} = j! \sum_{i=0}^{j}
\frac{\Gamma(b+i+1)}{\Gamma(b+1)}\;a^{j-i},
\end{equation}
which clearly proves the statement. Finally, since  the
eigenvalues of $\mathbf{R}_{N-1}$ are real, positive and
different, it follows that \textit{the eigenvalues of
$\mathbf{A}_{N-1}$ are real, negative and different}. Q.E.D.


\begin{thebibliography}{99}

\bibitem{kendall50}{D.G. Kendal, J. Roy. Statist. Soc., B
\textbf{12}, 278 (1950)}

\bibitem{bharucha53}{A.T. Bharucha-Reid, Biometrics, \textbf{9},
275 (1953)}

\bibitem{eigen71}{M. Eigen, Naturwissenschaften, \textbf{58}, 465
(1971)}

\bibitem{segre98}{D. Segr\'e, Y. Pilpel, D. Lancet, Physica A,
\textbf{249}, 558 (1998)}

\bibitem{ferreira03}{A.S. Ferreira, M.A. da Silva, and J.C.
Cressoni, Phys. Rev. Lett., \textbf{92}, 219901-1 (2004)}

\bibitem{lpal04}{L. P\'al, arXiv:cond-mat/0404402}

\bibitem{dean56}{P. Dean, Proc. Cambridge Phil. Soc., \textbf{52},
752 (1956)}

\bibitem{karlin58}{S. Karlin and J.L. McGregor, Techn. Rep.,
\textbf{9}, Stanford University (1958)}

\bibitem{bellman60}{R. Bellman, \textit{Introduction to matrix analysis},
McGraw-Hill Book Company, INC. New York Toronto London (1960)}

\bibitem{karlin57}{S. Karlin and J.L. McGregor, Trans. Amer. Math.
Soc., \textbf{85}, 489 (1957)}

\end{thebibliography}
\end{document}